\documentclass[prd,preprint,showpacs]{revtex4}
\usepackage{ booktabs}
\usepackage{stmaryrd}
\usepackage{mathrsfs}
\usepackage{amsmath}
\usepackage{latexsym}
\usepackage{amsfonts}
\usepackage{graphicx}
\usepackage{subfigure}

\def\be{\begin{equation}}
\def\ee{\end{equation}}
\def\ba{\begin{eqnarray}}
\def\ea{\end{eqnarray}}

\begin{document}
\title{Parameterization and Reconstruction of Quasi Static Universe}

\author{Jie Liu}
\email{liujie10b@mails.gucas.ac.cn}
\author{Yun-Song Piao}
\email{yspiao@gucas.ac.cn}

\affiliation{ College of Physical Sciences, Graduate University of
Chinese Academy of Sciences, Beijing 100049, China}

\begin{abstract}

We study a possible fate of universe, one in which there is
neither the a rip singularity, which results in the disintegration
of bound systems, nor an endless expansion, instead the universe
will be quasi-static. We discuss the parameterization of the
corresponding evolution and the reconstruction of the scalar field
model. We find, with the parameterization consistent with the
current observation, that the current universe might arrive at a
quasi-static phase after less than 20Gyr.

\end{abstract}
\pacs{98.80.-k, 95.36.+x}
 \maketitle

\section{Introduction}

The observations imply that the current universe is undergoing an
acceleration \cite{Riess}, which is driven by dark energy, or
lead by the modification of gravity on large scale. The simplest
candidate of dark energy is the cosmological constant, but it
suffers from the cosmic coincidence problem. Thus the dynamical
dark energy might be favored, there have been some candidates of
scalar field model such as
\cite{Wetterich},\cite{Gibbons},\cite{Kamionkowski},\cite{quintom},\cite{Wei},
see e.g.\cite{review1},\cite{review2} for reviews.

The fate of universe is determined by the nature of dark energy.
The universe driven by the phantom will evolve to a singularity,
in which the energy density become infinite at finite time, which
is called the big rip, see also Ref.\cite{NOT} for other future
singularities. How to avoid these singularities is still an
interesting issue, e.g.\cite{Nojiri1} and the little rip scenario
in which a rip singularity cannot occur in finite time
\cite{littlerip},\cite{littlerip1}.

In principle, due to the difference in the nature of dark energy, the
universe may have a different date. Here, we will study the 
possibility of the fate of universe, in which the universe will be
quasi-static some time after the current time. In this fate of
universe, there is neither the disintegration of bound systems,
i.e. the rip singularity, nor the moving apart stars and galaxies,
i.e. the endless expansion, e.g. in a dS universe or the universe
dominated by matter; instead the scale factor of universe will be
nearly constant.

The outline of paper is as follows. We will firstly study how to
parameterize the evolution of a quasi-static universe in section
II, and then will bring some specific cases in section III. The
reconstruction of the scalar field model is given in section IV.
A discussion is given in section V.


\section{How to parameterize the evolution of a quasi-static universe }

For the universe to be static, it is required that \be H\longrightarrow
0,\,\,\, and \,\,\, a \longrightarrow constant \label{Hrequire}\ee
for $t\longrightarrow \infty$ should be satisfied, where $H$ is
Hubble parameter and $a$ is the scale factor.
The corresponding evolutions can
be parameterized as \ba H & \sim & {1\over t_0}\left({ t_0\over
t}\right)^{b},\,\,\, \label{H1}\\ & or & Exp\left(-\left({t \over
t_0}\right)^k\right), \,\,\, \label{H2}\ea where $b>1$ and $k>0$
are required. The case with $0<b\leq 1$, e.g.$H\sim 1/t$, Eq.(\ref{Hrequire}) is not
suitable, since $\int H dt \longrightarrow 0$  diverges. Here,
the behavior of $H\longrightarrow 0$ is a power law or is
exponential. However, of course, the behavior of $H$ could be
also double exponential, \ba H & \sim & Exp\left(-e^{\left({t\over t_0}\right)}\right), \label{H3}\\
& or & \,\,{\rm a}\,\,{\rm higher}\,\, {\rm exponential}.\ea

Here, for (\ref{H1}), $a=e^{\int Hdt}$ is given by \ba a & \sim &
a_{static} Exp\left(\frac{1}{1-b}\left({ t_0\over t}\right)^{b-1}\right).
\label{model1}\ea Thus in the regime $t\gg t_0$, the universe is
asymptotically static, where $a_{static}$ is the static value of
$a$.

In this parameterization, the universe asymptotically arrives at
the static phase. However, it may be nearly static some time after
$t_0$. We define the time when e.g. $a\sim 0.99a_{static}$ as the
static time $t_{static}$, which means  the time that the universe
has become quasi-static. 
 With(\ref{model1}), $t_{static}$ is given by \be
t_{static}={ t_0\over [(1-b)ln{0.99}]^{1/ (b-1)}}. \ee We see that the
larger $b$ is, the earlier $t_{static}$ is, since the Hubble
parameter decays faster for larger $b$.

However, the expansion of the current universe is accelerated,
thus a consistent parameterization should not only overlap the
above parameterization for $t\gg t_0$, but also give the current
acceleration around $t\simeq t_0$.

\section{ Application to specific cases}


We will bring some specific parameterizations in this section, and
discuss the details of the models and restrict the parameters in
parameterizations with current observation.

\subsection{ The universe with parameterization (i)
 }
We have
\be
H(t)=\frac{A}{t_{0}}\Big(\frac{t}{t_{0}}\Big)^{k-1}Exp\Big(-B\Big(\frac{t}{t_{0}}\Big)^k\Big)
\label{mH}\ee
The model have three parameters $A$, $B$, and $k$, and obviously,
$A$, $B>0$ and $k>1$. With the Hubble parameter (\ref{mH}), $a$ is given
by \begin{equation} a(t)=Exp\Big(\int
H(t)dt\Big)=a_{static}e^{-\frac{A}{kB}Exp\biggl[-B\Big(\frac{t}{t_{0}}\Big)^k\biggl]}.
 \label{ma}\end{equation}
The relationships between the three parameters and the current
$a_0$ and $H_0$ are
\begin{equation}
a_0=a_{static}e^{-\frac{A}{kB}Exp(-B)},
 \label{ma0}\end{equation}
and \begin{equation}
 H_{0}=\frac{Ae^{-B}}{t_{0}}.
\label{mh0}\end{equation}

The universe consists of  dark matter and  dark energy. We
have, for a FRW spacetime,
\begin{equation} \rho_{DE}=\frac{3}{\kappa^{2}} H^2-\rho_{DM},
 \label{mrhoDE}\end{equation}
\begin{equation}
p_{DE}=-\frac{2}{\kappa^2}\dot{H}-\frac{3}{\kappa^{2}} H^2,
 \label{mpDE}\end{equation}
where $\kappa^2=8\pi G$. The density of dark matter is given as
\begin{equation}
\rho_{DM}=\rho_{DM0}Exp\biggl[
-\frac{3A}{kB}\Big(e^{-B}-e^{-B(\frac{t}{t_{0}})^k}\Big)\biggl].
 \label{mrhoDM}\end{equation} Thus the required $\rho_{DE}$ and $p_{DE}$ are only determined
by the parameterization of $H$ or $a$. The equation of state
parameter of dark energy is $\omega_{DE}=p_{DE}/\rho_{DE}$. In
infinite latetime, both $H$ and $\dot H$ tend to 0, we have
$\rho_{DE}\simeq -\rho_{DM}$ and $p_{DE}\simeq 0$. Thus at
infinite latetime, it is required that $ \rho_{DE}<0$, which just
sets off the positive density of dark matter, and
$\omega_{DE}\simeq 0$ is the same as that of dark matter, which
ensures that the decaying of their energy density with time are
same. Here, the dark energy may be the field or fluid, which
satisfies (\ref{mrhoDE}) and (\ref{mpDE}).
The derivation  of the Hubble parameter is
\begin{equation}
\dot{H}=\frac{A}{t_{0}^2}\Big(\frac{t}{t_{0}}\Big)^{k-2}\biggl[k-1-Bk\Big(\frac{t}{t_{0}}\Big)^k \biggl]e^{-B(\frac{t}{t_{0}})^{k}}.
 \label{mdh}\end{equation}
The expansion of universe at $t_0$ is accelerated. The quasi-static phase implies that the universe must begin deceleration at
the time $t_{tr}>t_0$, in which ${\ddot a}=0$. With ${\ddot
a}/a={\dot H}+H^2$ and (\ref{mdh}), we get
\begin{equation}
k-1-k B\left({t_{tr}\over t_{0}}\right)^k+A\left({t_{tr}\over
t_{0}}\right)^k Exp\Big[-B\left({t_{tr}\over
t_{0}}\right)^k\Big]=0. \label{mttr}\end{equation} Thus we have the
condition
  \begin{equation}
\frac{1}{1-B}>k>{Ae^{-B}-1\over B-1}.
 \label{mk}\end{equation}
With (\ref{ma}), the universe has become quasi-static at $t_{static}$
 \begin{equation}
 t_{static}=\left(\frac{1}{B}ln \frac{-A}{kBln0.99}\right)^{1/k}t_{0}.
\label{mtst}
\end{equation}

\begin{figure}[!ht]
\subfigure[]{
\begin{minipage}[c]{0.3\textwidth}
\centering
\includegraphics[width=4cm]{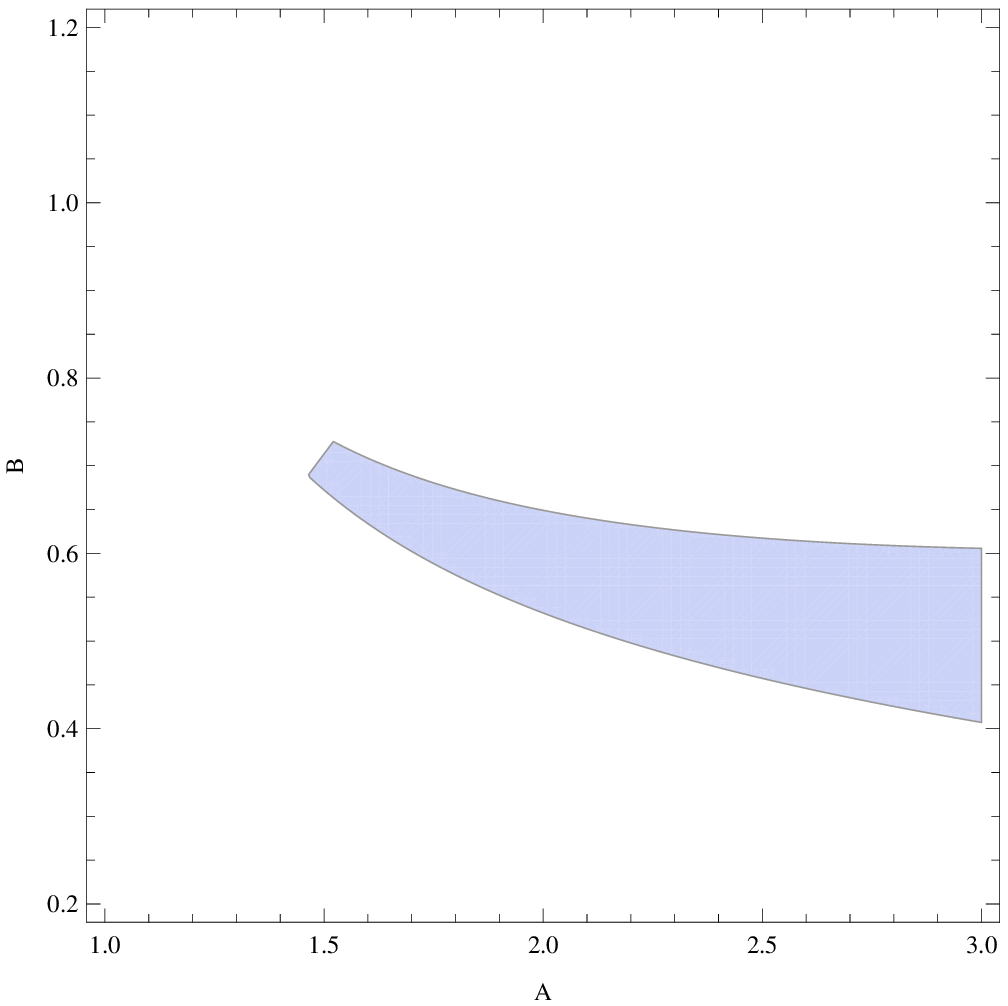}
\end{minipage}}%
\subfigure[]{
\begin{minipage}[c]{0.3\textwidth}
\centering
\includegraphics[width=4cm]{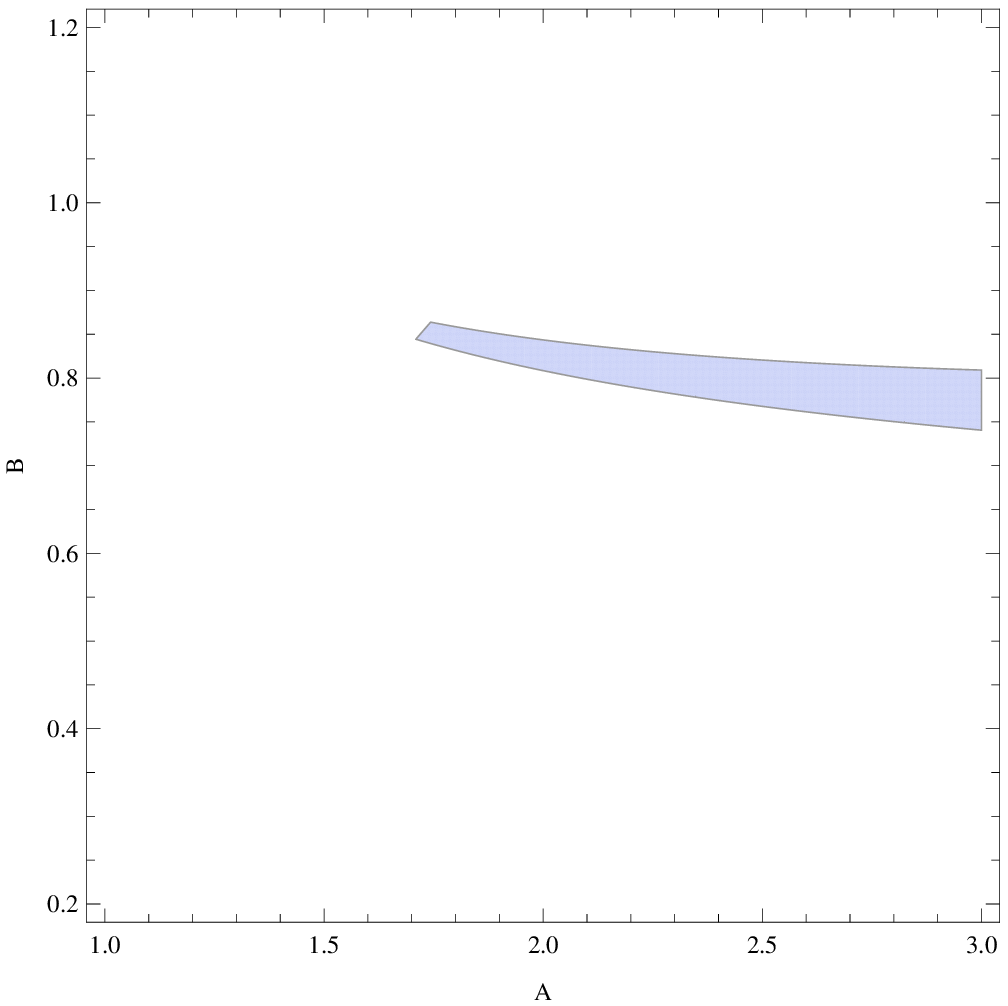}
\end{minipage}}%
\subfigure[]{
\begin{minipage}[c]{0.3\textwidth}
\centering
\includegraphics[width=4cm]{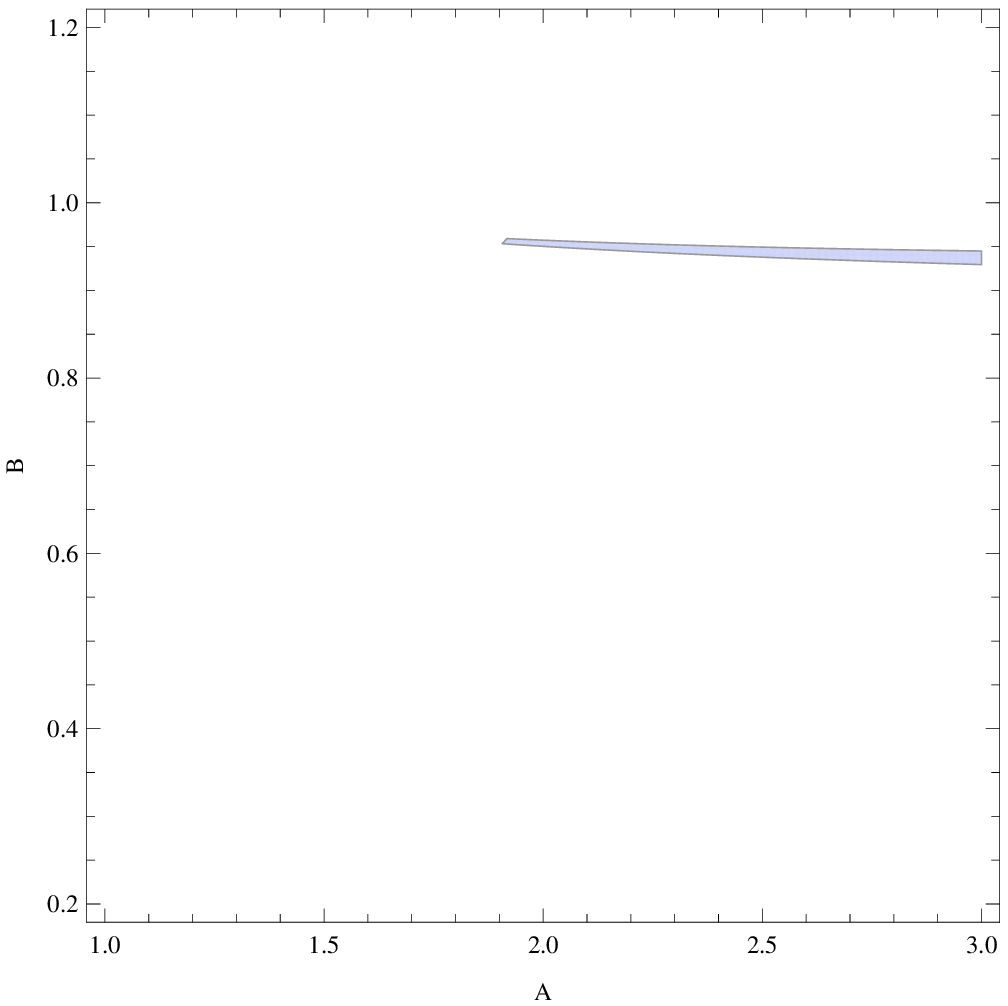}
\end{minipage}}
\caption{The observation constrains to the (A, B) parameters space
for the different  k.
 (a):~k=1.5,~~
 (b):~k=3,~~
 (c):~k=10. }
\label{fig1}
\end{figure}

We restrict the space of the parameters $A$ and $B$ with observation
in \cite{Komatsu}, in which $-1.033 \leq \omega_{DE} \leq -0.927$,
and the condition (\ref{mk}) for different $k$, see Fig.\ref{fig1}.
We find the range of the parameter space decrease with the increasing
$k$.


In Fig.\ref{fig2}, we plot the evolution of relevant quantities
numerically, in units of $\rho_{c0}^{-1}$, correspondingly the space of the parameters in
Fig.\ref{fig1}, where $\rho_{c0}=3H_{0}/\kappa^2$ is the present critical density of the universe. Here we take
$\Omega_{DM0}=0.27$. We find that the value of the parameter $A$ has a
range which is in agreement with the current observations,
i.e.$\Omega_{DE}\sim 3\Omega_{DM}$.

\begin{figure}
\setlength{\belowcaptionskip}{0pt}
\subfigure[]{
\begin{minipage}[c]{0.3\textwidth}
\centering
\includegraphics[width=4cm]{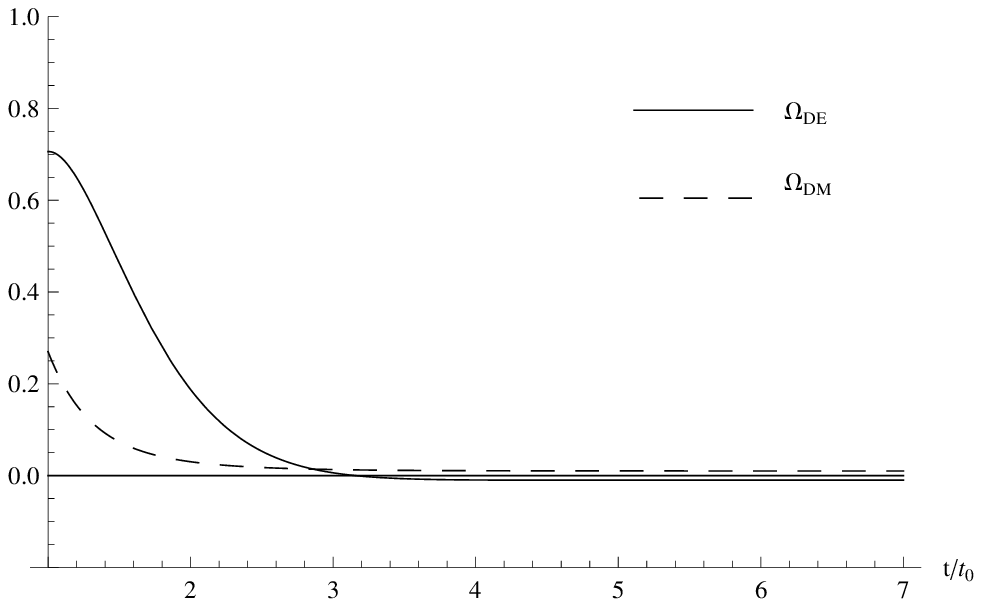}
\end{minipage}}%
\subfigure[]{
\begin{minipage}[c]{0.3\textwidth}
\centering
\includegraphics[width=4cm]{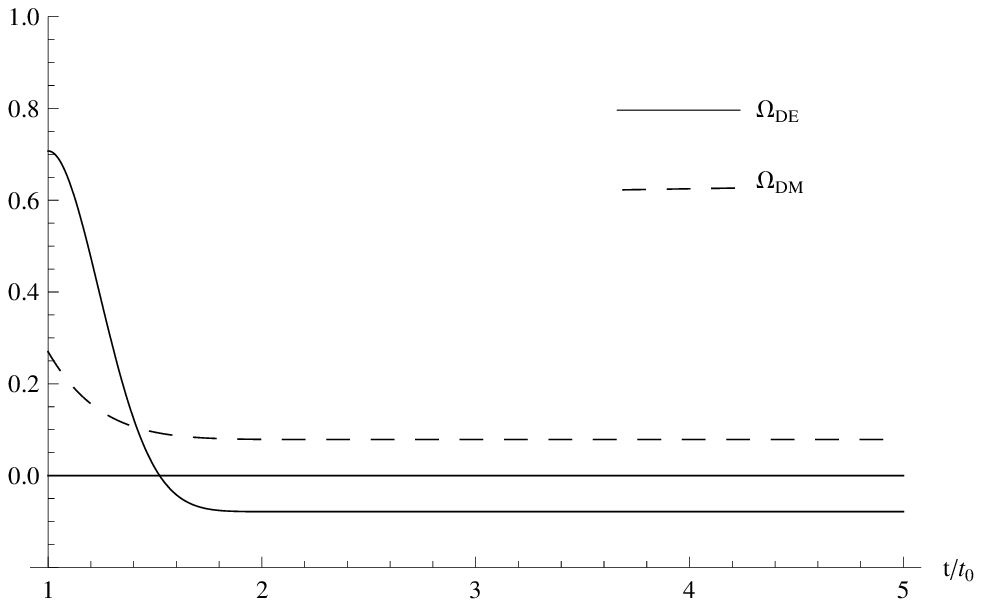}
\end{minipage}}%
\subfigure[]{
\begin{minipage}[c]{0.3\textwidth}
\centering
\includegraphics[width=4cm]{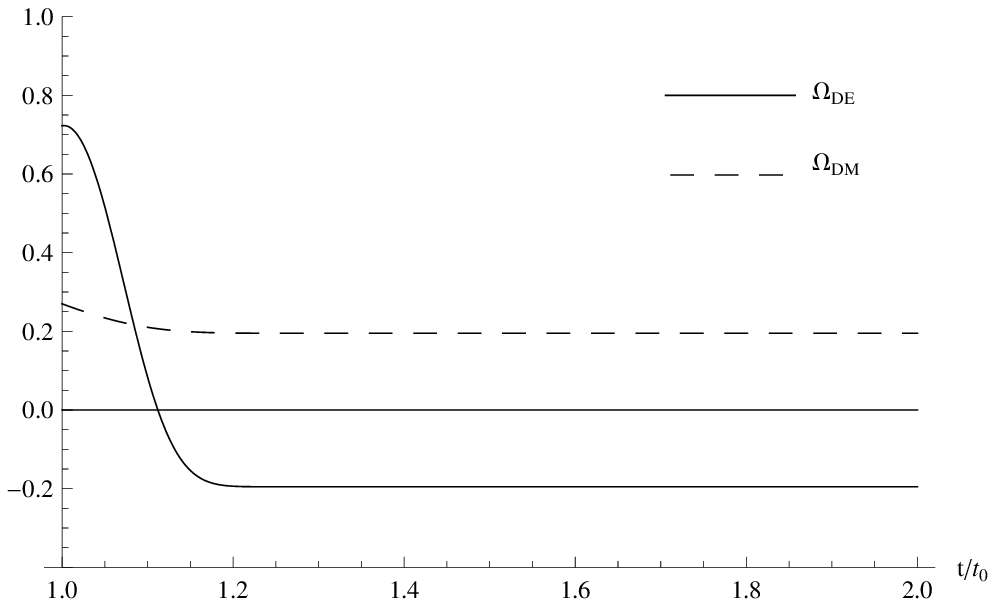}
\end{minipage}}
\label{fig2}
\end{figure}
\begin{figure}[!ht]
\setlength{\abovecaptionskip}{0pt}
\setlength{\belowcaptionskip}{0pt}
\subfigure[]{
\begin{minipage}[c]{0.3\textwidth}
\centering
\includegraphics[width=4cm]{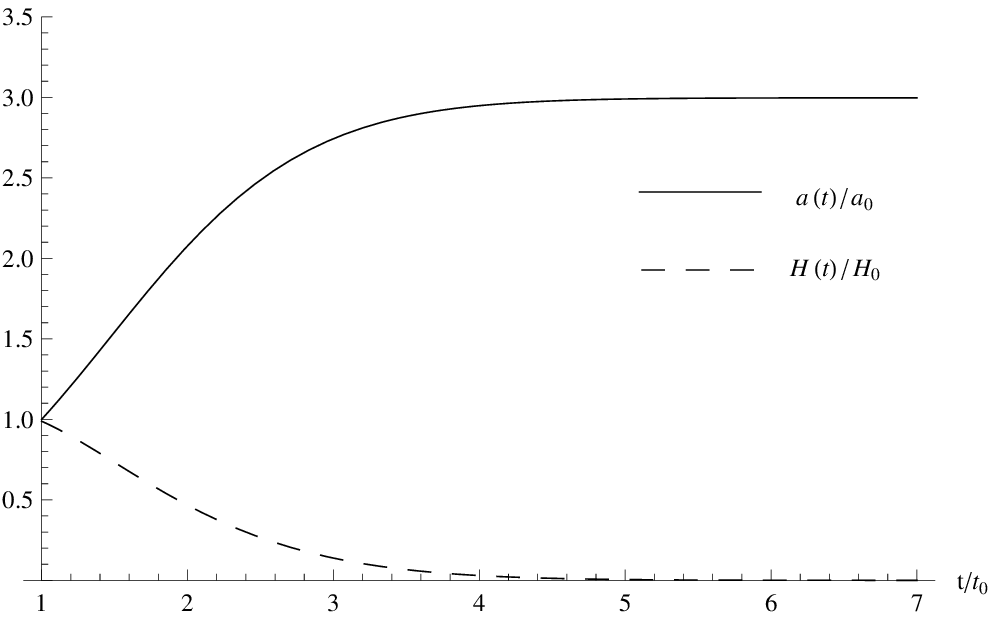}
\end{minipage}}%
\subfigure[]{
\begin{minipage}[c]{0.3\textwidth}
\centering
\includegraphics[width=4cm]{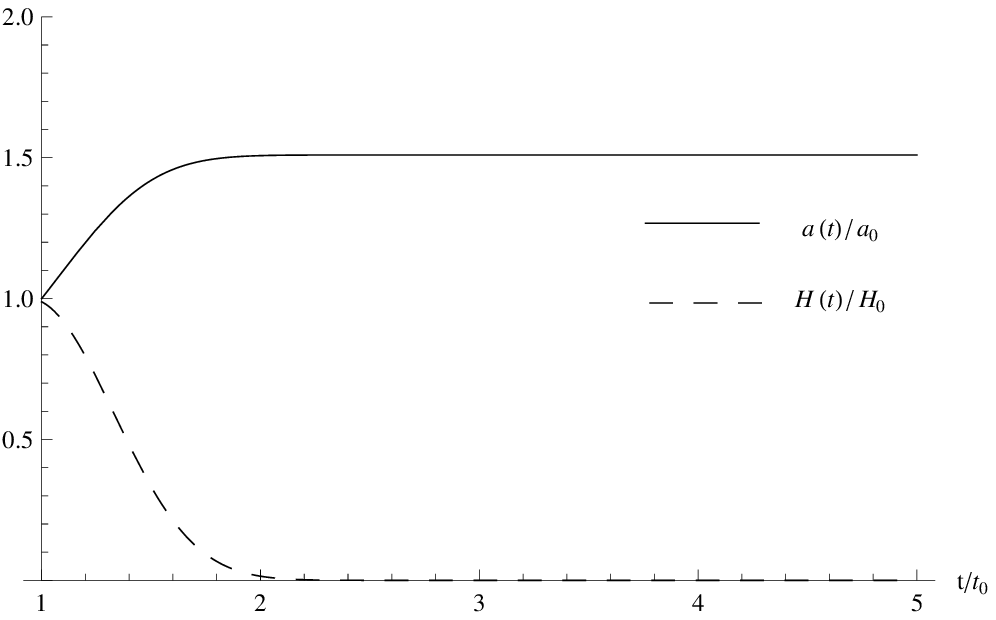}
\end{minipage}}%
\subfigure[]{
\begin{minipage}[c]{0.3\textwidth}
\centering
\includegraphics[width=4cm]{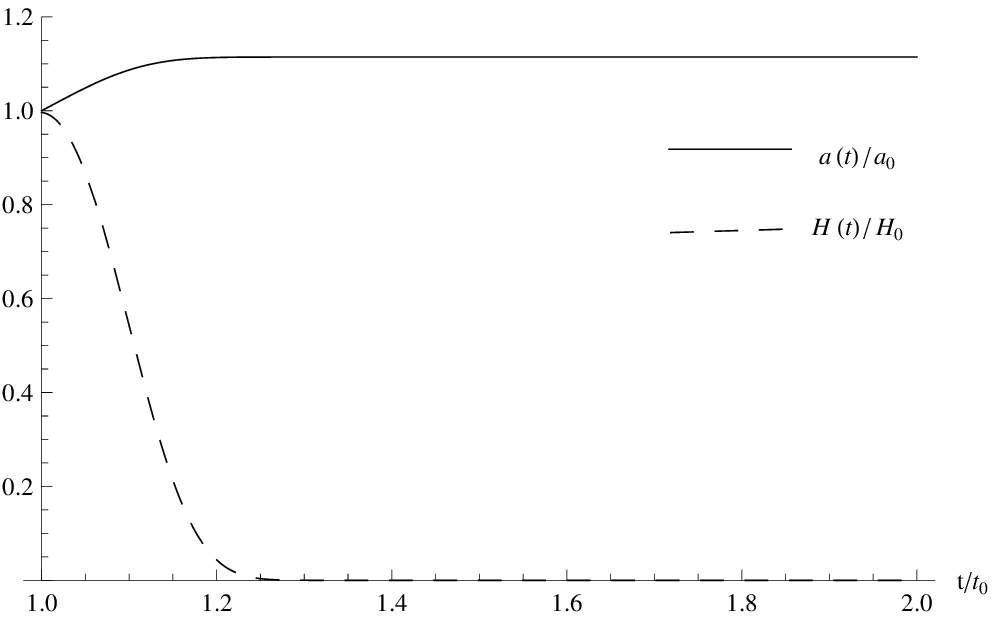}
\end{minipage}}
\caption{The upper panels are the graphs of the evolutions of
$\Omega_{DM}$ and $\Omega_{DE}$; the lower panels are the graphs
of the evolution of $a(t)$ and $H(t)$  by normalized.  (a) and (d):the used
parameter values are k=1.5,~ (A, B)=(1.8, 0.6);~~(b) and (e):the
used parameter values are ~k=3, ~(A, B)=(2.2, 0.8);~~(c) and
(f):the used parameter values are ~k=10, ~(A, B)=(2.5, 0.92). }
\label{fig2}
\end{figure}

In the Table I, we list
$t_{static}$ and $t_{tr}$ by choosing three sets of the special
parameters $A$ and $B$ for different $k$. We see that the larger
$k$ is, the earlier both $t_{tr}$ and $t_{static}$ are, since the
Hubble parameter decays faster for larger $k$.

\begingroup
\begin{table}[!ht]
    \label{spectrum}
    \begin{tabular}{|c|c|p{0.9cm}|c|c|c|}
       \hline \hline   k  &    (A, B)     &
       $t_{tr}/t_{0}$   &   $t_{tr}-t_{0}$ &  $t_{static}/t_{0}$  &  $t_{static}-t_{0}$
       \\ \hline \hline
      1.5  &    1.8,  0.6  &  1.468    & 64.116Gyr  &     4.270    & 447.990Gyr
      \\ \hline  \hline
      3    &    2.2,  0.8  &  1.079  & 10.823Gyr &   1.780 & 106.860Gyr
      \\  \hline \hline
      10   &    2.5,  0.92 &  1.008    &  1.096Gyr &     1.136 & 18.632Gyr
      \\ \hline \hline
    \end{tabular}
    \caption{The transition time $t_{tr}$ and the static time $t_{static}$
for the different parameter spaces.}
  \end{table}
  \endgroup

\subsection{ The universe with parameterization (ii)}
We have 
\begin{equation}
H(t)=\frac{A}{t_{0}}\Big(\frac{t_{0}}{t}\Big)^b\biggl[1-B\Big(\frac{t_{0}}{t}\Big)\biggl]
\label{2}
\end{equation}

\begin{figure}
\subfigure[]{
\begin{minipage}[c]{0.3\textwidth}
\centering
\includegraphics[width=4.4cm,,height=3cm]{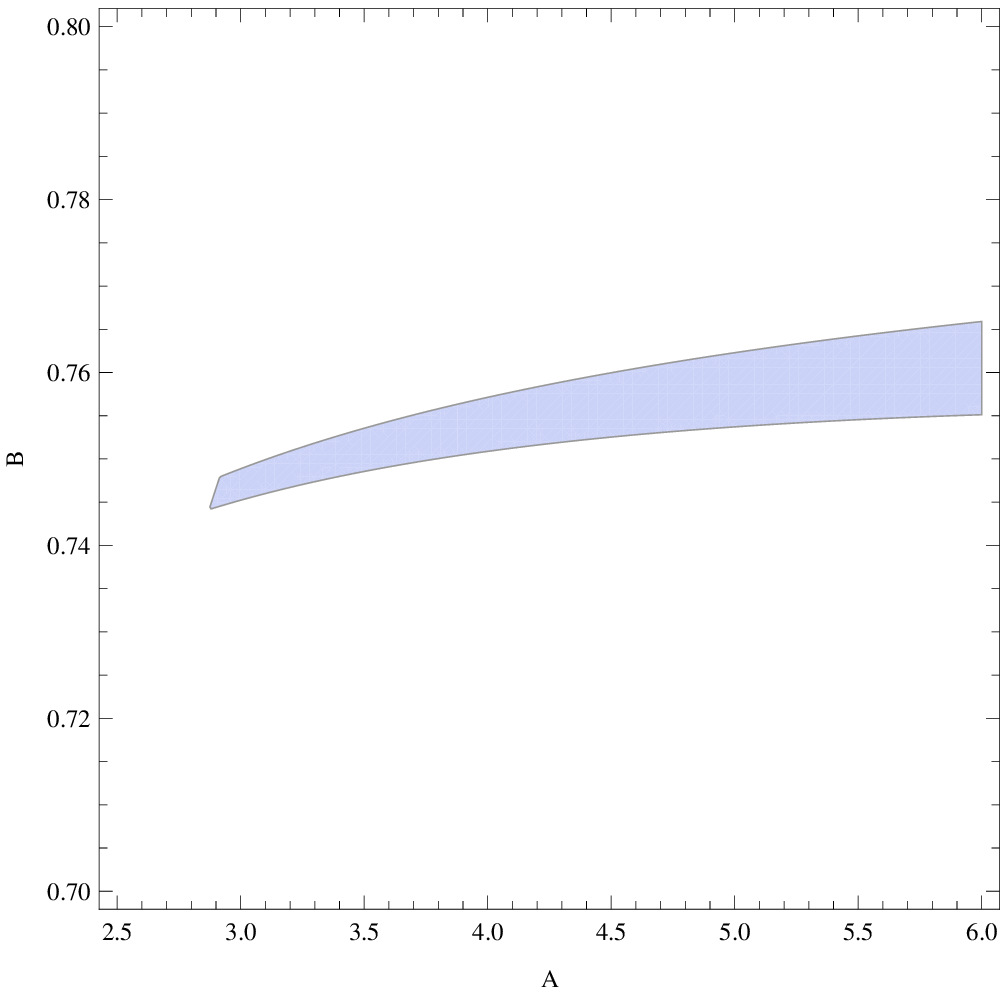}
\end{minipage}}%
\subfigure[]{
\begin{minipage}[c]{0.3\textwidth}
\centering
\includegraphics[width=4.6cm,height=3cm]{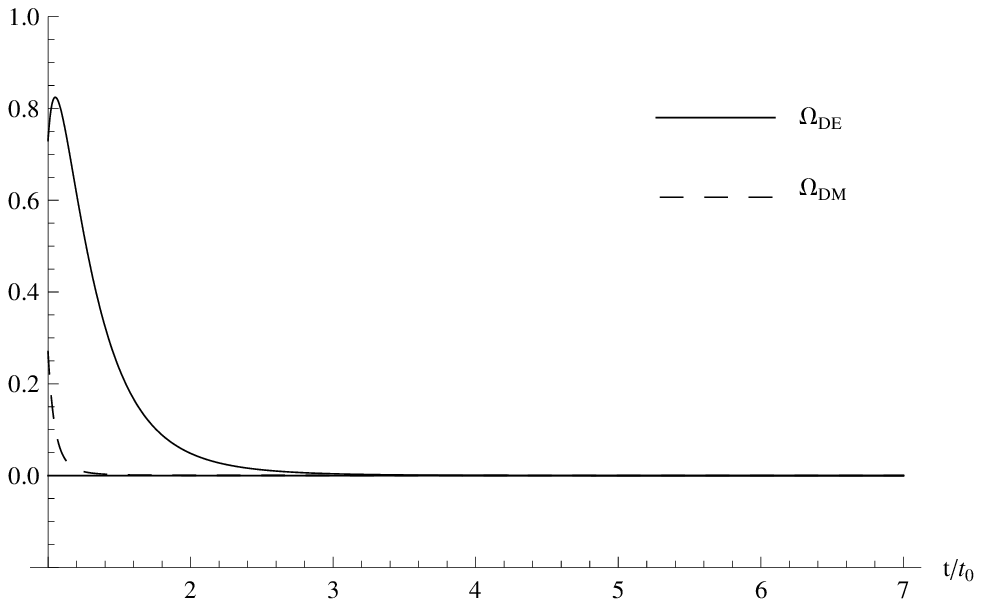}
\end{minipage}}%
\subfigure[]{
\begin{minipage}[c]{0.3\textwidth}
\centering
\includegraphics[width=4.6cm,,height=3cm]{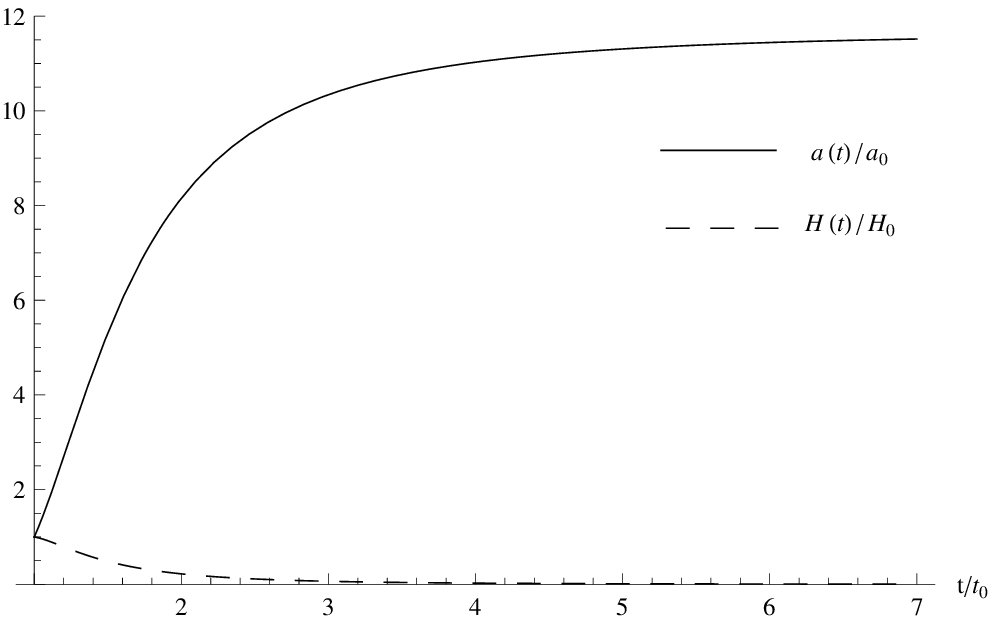}
\end{minipage}}
\caption{The left panel is the graph of the observation constrains
to the (A, B) parameters space for $b=3.5$; The middle panel is
the graph of the evolutions of $\Omega_{DM}$ and $\Omega_{DE}$;
The right panel is the graph of the evolutions of $a(t)$ and
$H(t)$. The used parameter values are $b=3.5$, $(A$, $B)=(4$,
$0.75)$} \label{fig3}
\end{figure}

The model have three parameters $A$, $B$, and $b$, and obviously,
$A$, $B>0$ and $b>2$. The discussion is similarly to the model in
subsection A. We will not repeat it, and only plot the evolution
of relevant quantities numerically in Fig.\ref{fig3} for a set of
parameters. The transition time is $t_{tr}=1.003t_{0}$, while the
static time is $t_{static}=7.804t_{0}$.



\section{The reconstruction of scalar field model}

We will reconstruct the scalar field model with the
parameterization (\ref{mH}). The reconstruction of scalar field
dark energy models has been studied in lots of references,
e.g.\cite{S1},\cite{Tsujikawa},\cite{Nojiri2},\cite{Guo},\cite{Zhangxin}.

The Lagrangian density of scalar field is given by $P(X,\phi)$,
where $X=-g^{\mu\nu}\partial_{\mu}\phi
\partial_{\nu}\phi/2$. Thus the energy density $\rho_{DE}=2XP_{X}-P$ is
derived. In a spatially flat FRW universe, the equations for the
dark matter and the scalar field are
\begin{equation}
\frac{3}{\kappa^{2}} H^2=\rho_{DM}+2X{P}_X- {P},
 \label{H}\end{equation}
\begin{equation}
\frac{2}{\kappa^2}\dot{H}=-\rho_{DM}-2X{P}_X,
 \label{DH}\end{equation}
where $\rho_{DM}=\rho_{DM0}({a_0\over a})^{3}$. Thus we have \be
P= -\frac{2}{\kappa^2}\dot{H}-\frac{3}{\kappa^{2}} H^2,
\label{P}\ee \be {\dot \phi}^2
P_X=-\frac{2}{\kappa^2}\dot{H}-\rho_{DM}. \label{dotphi}\ee

In principle, after specifying $P(X,\phi)$, we can have the
evolution of $\dot \phi$ and $P$ with the time. Thus the potential
function in $P(X,\phi)$ can be reconstructed.

\begin{figure}
\centering
\includegraphics[width=6.8cm,height=6.5cm]{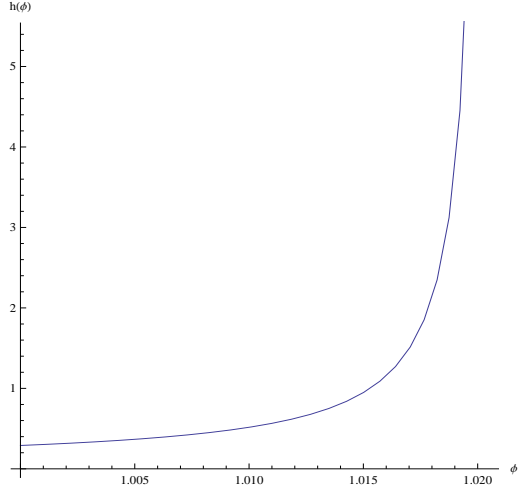}
\caption{  The reconstruction for the function  of  $h(\phi)$ according to the model A with $k=10$ and $(A$, $B)=(2.5$, $0.92)$.}
 \label{fig4}
\end{figure}

Here, we apply a generalized ghost condensate Lagrangian,
e.g.\cite{Tsujikawa}, in which $P=-X + h(\phi)X^2$. Thus with
Eqs.(\ref{P}) and (\ref{dotphi}), we have \be {\dot \phi}^2=
\frac{6}{\kappa^2}\dot{H}+\frac{12}{\kappa^{2}}
H^2-\rho_{DM},\label{hh}\ee


\be \begin{split}h(\phi)~ =~
{\frac{4}{\kappa^2}\dot{H}+\frac{12}{\kappa^{2}} H^2-2\rho_{DM}\over
(\frac{6}{\kappa^2}\dot{H}+\frac{12}{\kappa^{2}} H^2-\rho_{DM})^2 }
~=~\kappa^2t^2\times~~~~~~~~~~~~~~~~~~~~~~~~~~~~~~~~~~~~~~~~~~~~~~~~~~~~~~~~~~~~~~~~~~~~~~~~~~\\
\frac{4Ae^{-B(t/t_{0})^k}\Big(\frac{t}{t_{0}}\Big)^k\Big[(3Ae^{-B(t/t_{0})^k}-kB)\Big(\frac{t}{t_{0}}\Big)^k-1+k\Big]-2Exp\Big[-\frac{3A}{kB}\Big(e^{-B}-e^{-B(t/t_{0})^k}\Big)\Big]\kappa^2t^2\rho_{DM0}
}{\Big\{6
Ae^{-B(t/t_{0})^k}\Big(\frac{t}{t_{0}}\Big)^k\Big[(-2Ae^{-B(t/t_{0})^k}+kB)\Big(\frac{t}{t_{0}}\Big)^k+1-k\Big]+Exp\Big[-\frac{3A}{kB}\Big(e^{-B}-e^{-B(t/t_{0})^k}\Big)\Big]\kappa^2t^2\rho_{DM0}\Big\}^2}
. \end{split} \label{h}\ee
 With (\ref{hh}) and (\ref{h}), one can reconstruct the
function  of $h(\phi)$  for the generalized ghost condensate model
in the quasi static universe. Considering that both $h(\phi)$ and
$\phi$ are parametric equations of $t$, we can plot the
function graph of the h($\phi$) over $\phi$  numerically.  In
Fig.{\ref{fig4}}, we plot  the reconstruction for  $h(\phi)$
according to  with the parameterization (\ref{mH})  numerically, in
units of $\rho^{-1}_{c0}$,  and the special  parameters are $k=10$
and $(A$, $B)=(2.5$, $0.92)$. Here, we have fixed the field
amplitude at the present epoch to be 1: $\phi(t_{0})=1$.

Let us check the stability of the system.  $hX = 1/2$ corresponds to
the cosmological-constant model, and  the system can enter the
phantom region when ($hX < 1/2$), so the model could be unstable  as
a phantom. As has been pointed out by Tsujikawa
\cite{Tsujikawa,Tsujikawa1}, when considering the stability of
classical perturbations, two  quantity are usually taken into
account : $\xi_{1} =P_{X}+2XP_{XX}\geq0$, and $\xi_{2}=P_{X}\geq0$.
Combining Eq.(\ref{h}), we need $ 2\dot{H}+3H^2>0$ so that the
perturbations are classically stable. The conditions is easy to
satisfy for the model (i) with the special  parameters: $k=10$ and
$(A$, $B)=(2.5$, $0.92)$. For some case,  $P_{X}/(P_{X}+2XP_{XX})$
will become negative. This instability may be avoided if the phantom
behavior is just transient(see Ref. \cite{Tsujikawa1}).

\section{Discussion}

The fate of the universe is an interesting issue. We study a
possibility of the fate of the universe, in which the universe has become quasi-static some time after $t_0$. This model may has a region of
parameter space in which it resembles $\Lambda$CDM, which thus is
not conflict with the current observation.

In section III.A, we study the parameterization (\ref{mH}). When
$t>t_0$, $H(t)$ is rapidly decreased, thus we will get into the
static universe shortly after $t_0$. We find that the current
universe might  arrive at the static phase after about
18.632 Gyr. For the universe, it is a very brief spell.

In section III.B, we study the parameterization (\ref{2}). Here,
since $H(t)$ is decreased slower than that in section III.A, it
will take a longer time to get into the static universe. With specific
parameterization consistent with the current observation, we find
that the current universe will arrive at the static phase after
 $t_{static}=7.804t_{0}$, about 932.148 Gyr.

Here, we show the possibility of an alternative fate of universe,
in which the universe has become quasi-static some time after $t_0$. In
this fate of universe, there is neither the disaggregation of
bound systems, nor the moving apart of stars and galaxies, the universe
will have a quiet ``afternoon".

The reconstruction of the scalar field models of quasi-static
universe is significant. We have discussed a case. However, we
have neglected the coupling between the scalar field and the dark
matter. The case including the coupling is also interesting; it
will be a substantial work.

\textbf{Acknowledgments}

We thank Hong Li, Taotao Qiu and Yuanzhong Zhang for helpful discussions. This work is
supported in part by NSFC under Grant No:11075205, in part by the
Scientific Research Fund of GUCAS(NO:055101BM03), in part by
National Basic Research Program of China, No:2010CB832804

\end{document}